\documentclass[a4paper,fleqn]{cas-sc}

\usepackage{natbib}
\usepackage{amsthm}
\usepackage{todonotes} 
\usepackage{graphicx}
\usepackage{gensymb}
\usepackage{algorithm}
\usepackage{algpseudocode}
\graphicspath{ {./images/} }
\usepackage{esvect}
\usepackage{nicefrac}
\usepackage{hyperref}
\usepackage{ulem}
\usepackage[export]{adjustbox}
\usepackage{lineno}

\def\tsc#1{\csdef{#1}{\textsc{\lowercase{#1}}\xspace}}
\tsc{WGM}
\tsc{QE}
\tsc{EP}
\tsc{PMS}
\tsc{BEC}
\tsc{DE}

\newcommand*{\figref}[2][]{%
  \hyperref[{fig:#2}]{%
    Fig.~\ref*{fig:#2}%
    \ifx\\#1\\%
    \else
      .#1%
    \fi
  }%
}

\newcommand*{\tableref}[2][]{%
  \hyperref[{tab:#2}]{%
    Tab.~\ref*{tab:#2}%
    \ifx\\#1\\%
    \else
      .#1%
    \fi
  }%
}

\newcommand*{\eqnref}[2][]{%
  \hyperref[{eq:#2}]{%
    Eqn.~\ref*{eq:#2}%
    \ifx\\#1\\%
    \else
      .#1%
    \fi
  }%
}

\newproof{pf}{Proof}

\shorttitle{SSPD - Advanced Scintillator-SiPM Particle Detector} 

\shortauthors{Simhony Yoav et~al.}

\title [mode = title]{Scintillator-SiPM Detector for Tracking and Energy Deposition Measurements}

\author[1,2]{Simhony Yoav}[type=editor]
\ead{yoavsimhony@mail.tau.ac.il}

\author[3] {Segal Alex}
\ead{alexs@afeka.ac.il}

\author [1] {Orlov Yuri}
\ead{ureor1@gmail.com}

\author[2]
{Amrani Ofer}
\ead{ofera@tauex.tau.ac.il}

\author[1,4]
{Etzion Erez}
\ead{ereze@tauex.tau.ac.il}

\affiliation[1]{organization={Tel Aviv University},
    addressline={Raymond and Beverly Sackler School of Physics and Astronomy}, 
    city={Tel Aviv},
    postcode={69978}, 
    country={Israel}}
    
\affiliation[2]{organization={Tel Aviv University},
    addressline={School of Electrical Engineering}, 
    city={Tel aviv},
    postcode={69978}, 
    country={Israel}}

\affiliation[3]{organization={Afeka College of Engineering},
    addressline={Unit of Mathematics}, 
    city={Tel Aviv},
    postcode={6910721}, 
    country={Israel}}

\affiliation[4]{organization={University of Toronto},
    addressline={Department of Physics}, 
    city={Toronto},
    postcode={M5S 1A2}, 
    state={ON},
    country={Canada}}    

\begin{abstract}
An innovative particle detector that offers a compelling combination of cost-effectiveness and high accuracy is introduced. The detector features plastic scintillators paired with a sparse arrangement of SiPMs, strategically positioned within a unique opto-mechanical framework. This configuration delivers precise measurements of spatial impact position and energy deposition of impinging particles.
The manuscript describes the detector's physical model complemented by an analytical representation. These calculations underpin a numerical algorithm, facilitating the estimation of particle impingement position and energy deposition. The results of the numerical calculations are compared with the output of GEANT4 simulations and evaluated by rigorous laboratory testing. An array of these detectors, intended for deployment in a spaceborne experiment, underwent detailed design, manufacturing, and testing. Their performance and alignment with the physical model were validated through meticulously conducted ground-based laboratory experiments, conclusively affirming the detector's properties.
\end{abstract}

\begin{document}


\maketitle
\section{Introduction}

We present a new detector, the Scintillator-SiPM Particle Detector (SSPD), which integrates scintillators and Silicon Photomultipliers (SiPMs). The SSPD offers the capability to reconstruct a traversing particle's 2D spatial position and its energy loss within the detector.

This detector design and its accompanying physical model demonstrate enhanced characterization of particles traversing the detector compared to prior work~\cite{heredge2021muon, cashion2021muon}. The authors initially detailed this detector in~\cite{simhony2021tausat} and subsequently tested it aboard the International Space Station (ISS)~\cite{COTS-Capsule-System}. 

This paper outlines the guidelines for the detector design, followed by details of implementation, manufacturing, and testing. Additionally, the paper provides an approximate physical model and a more accurate physical model, along with formulas and algorithms for deriving impinging particle localization and energy deposition estimation. These algorithms were compared with GEANT4~\cite{GEANT4:2002zbu} simulation results and verified by laboratory-conducted secondary cosmic ray measurements with the COTS-Capsule hodoscope~\cite{COTS-Capsule-System}. A comparison of different simulations and measurements for both models is presented.

\section{SSPD Design} \label{Design and Physical Model of the SSPD}

\begin{figure}[htbp]
\centering
\includegraphics[width=0.7\textwidth]{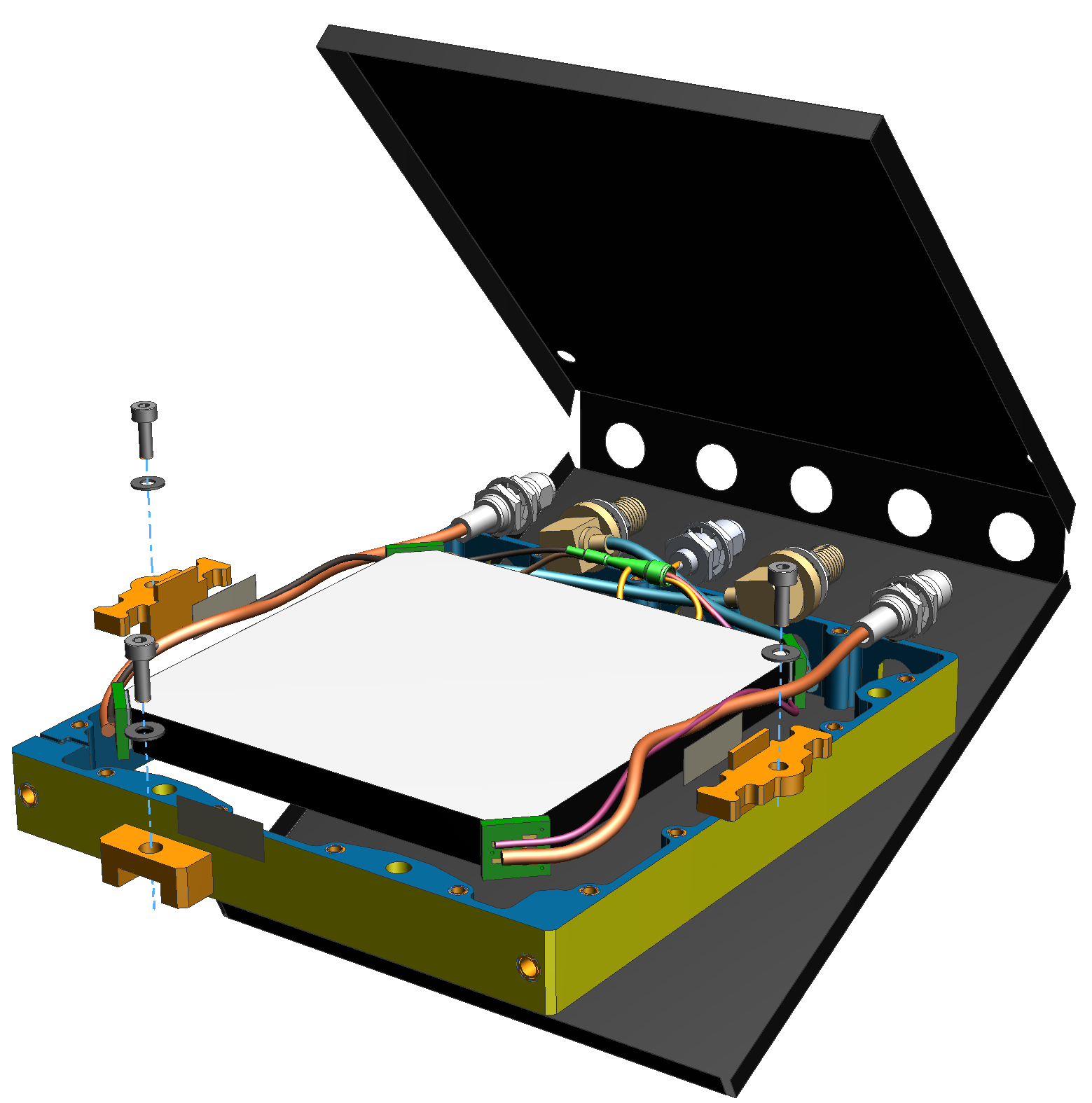}
\caption{CAD design of a COTS-Capsule SSPD. Conspicuously depicted is a yellow aluminum frame equipped with five SMA connectors. Four SensL FJ-60035 SiPM boards, colored green, are optically coupled to the truncated corners of an EJ-200 prism-shaped scintillator and electrically connected to the SMA connectors. Also visible are the black-coated side faces of the scintillator. This SSPD is subsequently sandwiched by a folded sheet of Thorlabs BKF12 matte black $70$ $\mu m$ aluminum foil.}
  \label{fig:3D_model_scintillator_detector}
\end{figure}

The SSPD discussed herein is a prism-shaped bulk plastic scintillator with optically coupled SiPM sensors, as illustrated in \figref{3D_model_scintillator_detector}. To mitigate stray light, certain faces of the scintillator are coated with a light-absorbing material, while other faces remain uncoated, maintaining a pristine surface finish such as polished or as-cast. The assembly is housed within a light-tight enclosure to minimize external light interference.

The SSPD was specifically designed as part of the COTS-Capsule spaceborne experiment~\cite{COTS-Capsule-System} and launched and tested aboard the ISS. The outer dimensions of the scintillator bulk measure $70  \times 70  \times 6.7~$ mm$^3$ where the $z$ dimension is along the shorter axis. The scintillator's four corners are truncated at a $45 \degree $ angle, resulting in four surfaces measuring $6.7  \times 6.7~$ mm$^2$, each optically coupled with a SiPM sensor. This truncation directs the SiPMs toward the scintillator's center.  

To eliminate stray light within the scintillator, black paint is applied to the four longer side faces ($70 \times 6.7~$ mm$^2$). This step is crucial for accurate operation, allowing precise estimation of impinging particle position and energy. Additionally, the top and bottom surfaces 
maintain a pristine surface finish, essential for effectively channeling light within the scintillator, thus enhancing its yield and sensitivity. The entire setup is enclosed within a light-absorbing and light-tight enclosure, ensuring minimal interference from stray light. The scintillator is suspended "in midair" within this enclosure, with the top and bottom surfaces left undisturbed.

When a particle traverses the scintillator, it excites the material, leading to the emission of photons in all directions along the particle track. The SSPD's optical system is designed to guide this emitted light within the scintillator through total internal reflection from the top and bottom surfaces. Light that reaches the SiPMs is detected, while stray light is almost entirely absorbed, thereby not influencing the SiPMs' measurements.

\begin{figure*}
	\centering
		\includegraphics[width=\textwidth]{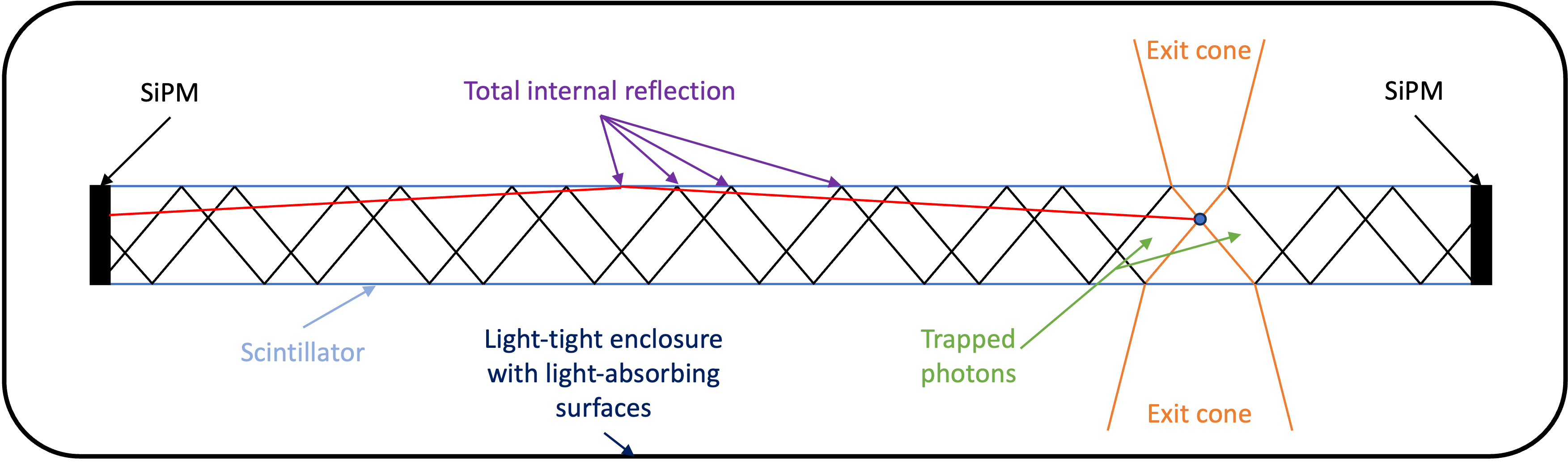}\label{fig:figure3}
\caption{Side cut view illustrating the internal structure of an SSPD. Photons originating from the position marked by the blue dot may travel in one of two distinct directions: within the "Exit cones" (depicted in orange), exiting either the top or bottom surface of the scintillator or in the direction of "Trapped photons" (illustrated by the black and red photon tracks), where they undergo multiple total internal reflections within the scintillator. Depending on their trajectory, trapped photons may either reach one of the SiPMs or be absorbed by the light-absorbing faces of the scintillator or the enclosure.}
  \label{fig:scintillator_detector_snell_law}
\end{figure*}

As illustrated in \figref{scintillator_detector_snell_law}, photons originating from the blue dot position and traveling within the "Exit cone" (orange) will exit the top or bottom surface of the scintillator and subsequently be absorbed by the enclosure's internal light-absorbing surfaces. Conversely, photons originating from the blue dot position and traveling within the "Trapped photons" direction, undergo multiple "total internal reflections" from the scintillator's top and bottom surfaces. These photons travel within the scintillator until they reach a SiPM sensor or encounter one of the scintillator's four light-absorbing surfaces.

The signals acquired from the SiPM sensors allow the SSPD to achieve precise 2D (x,y) localization of particle impact position and accurate estimation of the deposited energy within the scintillator.

\section{Physical Model Approximations}

At a macro view, the scintillator's waveguide attribute causes light to behave more like ripples spreading across the surface of a pond than the typical three-dimensional spherical radiation originating from a point source in space. This attribute affects the number of photons reaching each SiPM. This number is determined by the following: the total photon emission from the "point-source"; the (2D) angle ($\alpha_{ij}$ as depicted in \figref{2D_model_scintillator_detector_angles_to_sensors}); the ratio of refractive indices between the scintillator and its surroundings which define the "exit cones'" and "trapped photons'" angles.

Determining the particle's impinging position within the detector and measuring the deposited energy bears resemblance to the principles governing Global Navigation Satellite Systems (GNSS)~\cite{milliken1978principle} in their determination of position and time. Although the specific mathematical formulations differ, the fundamental concepts are similar.

\subsection{Perpendicularly Impinging Particle Model}

\begin{figure}[htbp]
\centering
\includegraphics[width=0.7\textwidth]{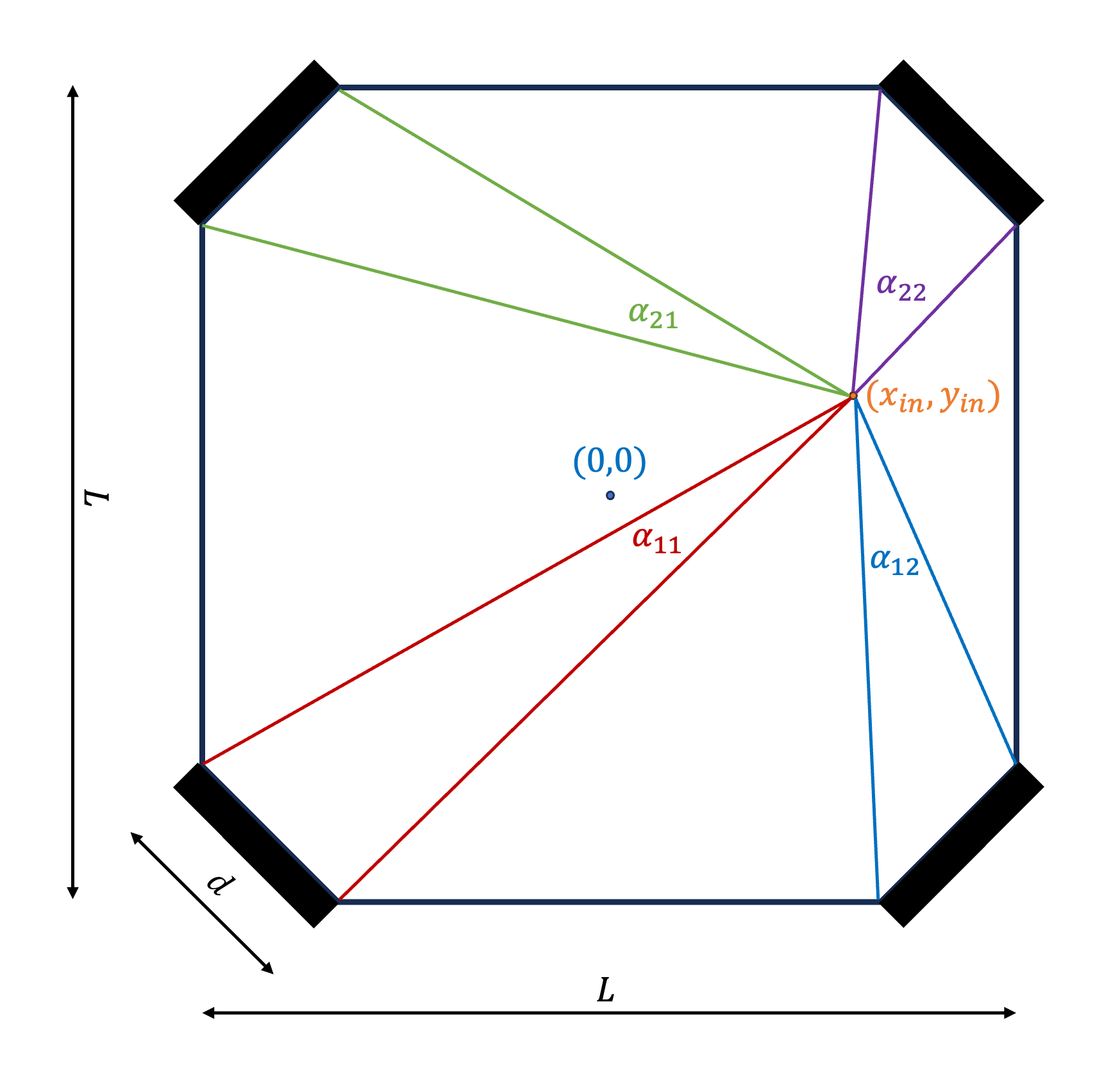}\label{fig:figure4}
\caption{Top view of an SSPD with the impinging position of a perpendicularly impinging particle $(x_{in},y_{in})$ (not-to-scale). Thick black lines at the corners indicate SiPM positions. The angles formed between the edges of the SiPM's active area and the impinging particle position within the scintillator are denoted by $(\alpha_{11},\alpha_{12},\alpha_{21},\alpha_{22})$.}
  \label{fig:2D_model_scintillator_detector_angles_to_sensors}
\end{figure}

\figref{2D_model_scintillator_detector_angles_to_sensors} illustrates a top view of an SSPD, showing the impinging position of a particle traveling perpendicularly to the surface, denoted as $(x_{in}, y_{in})$. The number of photons reaching each SiPM is proportional to the angles represented by $(\alpha_{11}, \alpha_{12}, \alpha_{21}, \alpha_{22})$.

Under the assumption of a particle traversing the scintillator perpendicularly, the following analytical equations hold:

\begin{equation}
N_{ij} = C_s \cdot d\cdot \textrm{LET}\cdot \eta_{\text{SiPM}_{ij}} \cdot \frac{\alpha_{ij}}{2\pi}, \qquad 1\leq i,j\leq 2.
\label{eq:detector_formulas_simple}
\end{equation}
where 
\[
C_s = \rho \cdot \eta_{\text{scint}}(\textrm{LET}, particle\; type) \cdot \sqrt{1-\frac{1}{n_{scint}^2}}.
\]

\begin{figure}[htbp]
\centering
\includegraphics[width=0.7\textwidth]{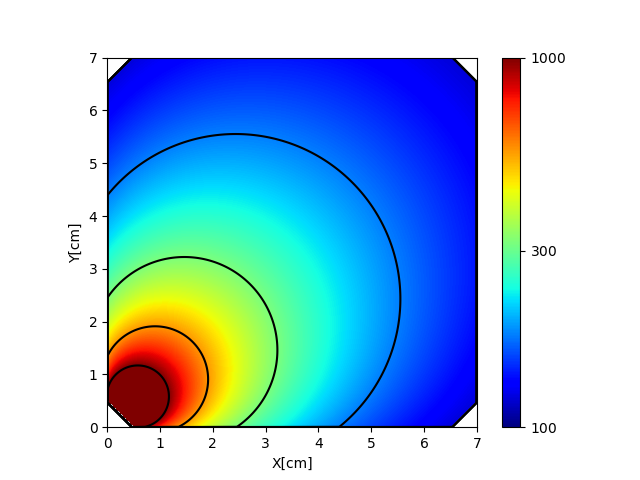}
\caption{2D heatmap showing simulation results of an SSPD illustrating the number of photons arriving at the bottom left corner SiPM vs. the particle's impinging position. Additionally, the simulation depicts contours of equal numbers of photons arriving at the SiPM, corresponding to \(N_{ph}=10^2, 10^{2.25}, 10^{2.5}, 10^{2.75}, 10^3\).}
\label{fig:2D_model_scintillator_detector_contour_lines}
\end{figure}

\noindent
\begin{itemize}
\item \(N_{ij}\) – Number of photons detected by SiPM$_{ij}$. 
\item \(\rho\) – Scintillator density; $n_{\text{scint}}$ – Refractive index; $\eta_{\text{scint}}(\textrm{LET}, particle\; type)$ – Scintillator efficiency for converting the energy deposited by the particle to irradiated photons~\cite{Workman:2022ynf} (For an EJ-200 scintillator, their values  are
$\rho \approx1.023
~$g/cm$^3$, $n_{\text{scint}} \approx 1.58$, $\eta_{\text{scint}} \approx 10,000~$ photons/MeV per $e^-$ ~\cite{eljen_scintillator}).   
\item \(d\) – Scintillator's width (0.67~cm). 
\item \(L\) – Scintillator height and length (7~cm). 
\item \(\eta_{\text{SiPM}_{ij}}\) – Efficiency of SiPM at corner ij (SiPM's active area divided by the scintillator's truncated corner's area, SiPM's Quantum efficiency, and bonding clear epoxy transparency). 
\item \((x, y)\) – Particle's impinging position at scintillator's half-width (\(d/2\)). 
\item \text{LET} – Particle's "Linear Energy Transfer" in $\left[\frac{\mathrm{MeV~cm^2}}{\mathrm{mg}}\right]$.
\end{itemize}
Visualizing these equations for a single particle with a LET of 1~$\frac{\mathrm{MeV~cm^2}}{\mathrm{mg}}$, 
perpendicularly impinging an Eljen EJ-200 scintillator, is illustrated in the 2D heatmap shown in \figref{2D_model_scintillator_detector_contour_lines}. 

The figure shows an inverse relationship between the number of photons detected by the bottom left SiPM sensor and the distance from the SiPM to the impinging particle position. Each SiPM exhibits equi-intensity contours, indicating that two events with identical intensities along these lines produce the same signal for a given SiPM sensor.

When data from two SiPMs are combined, two such contours intersect at only one or two specific $(x, y)$ locations. With three SiPMs in this geometry, we can determine both the 2D impinging particle location and the total number of photons released during the event. This corresponds to the energy deposited by the particle interacting with the SSPD.

For the physical model described, these equations accurately represent the number of photons reaching each SiPM for the entire scintillator area, except particles impinging the scintillator extremely close to the SiPMs. In such cases, some photons manage to reach and be detected by the SiPMs, even if their trajectories are directed within the exit cones. 

\subsection{Non-Perpendicularly Impinging Particle Model}

\begin{figure}[htbp]
    \centering
    \begin{minipage}[t]{0.7\textwidth}
        \centering
        \includegraphics[width=\linewidth]{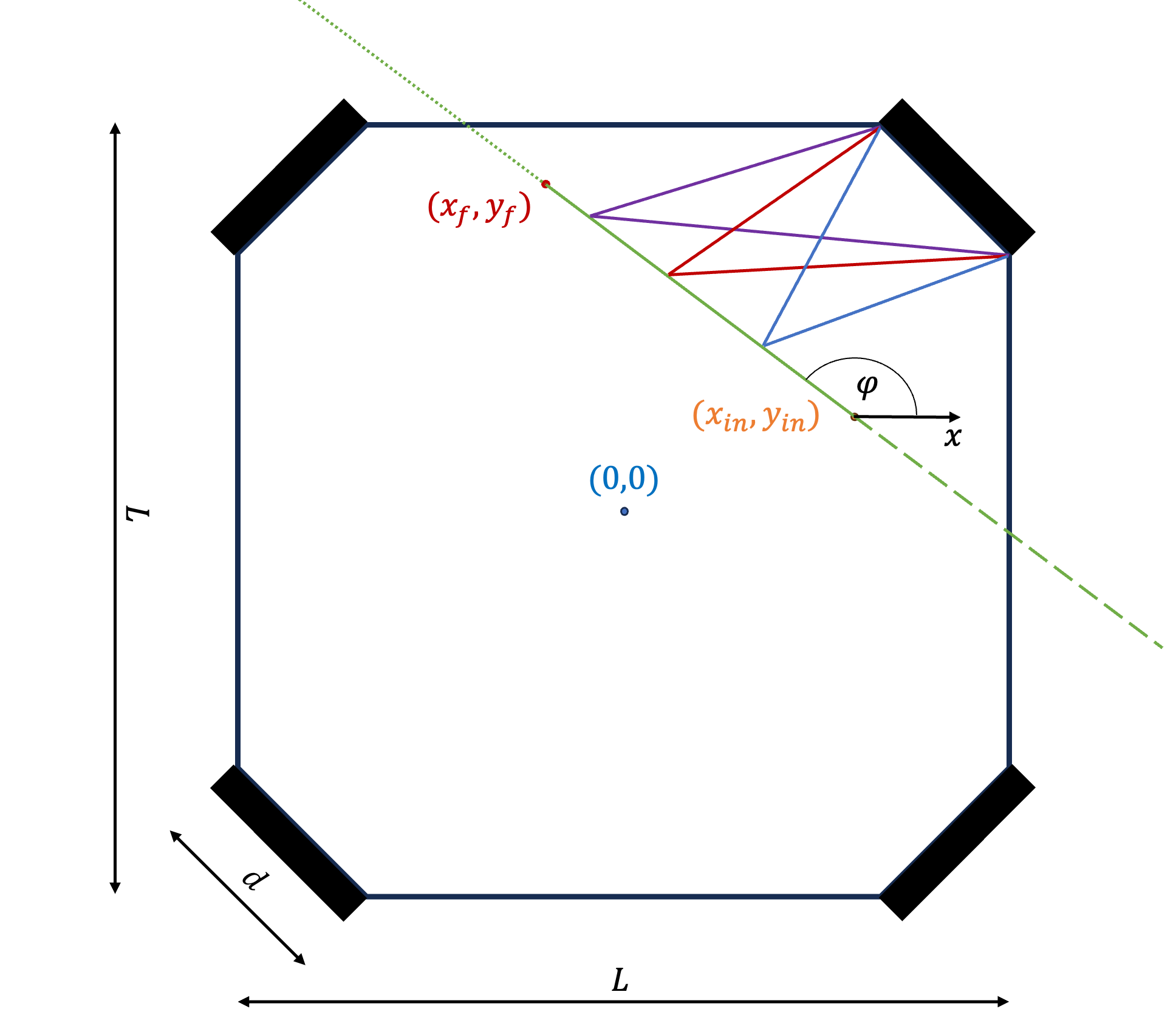}
        (a)
    \end{minipage}\\
    \hspace{1.8cm}\begin{minipage}[t]{0.6\textwidth}
        \includegraphics[width=\linewidth]{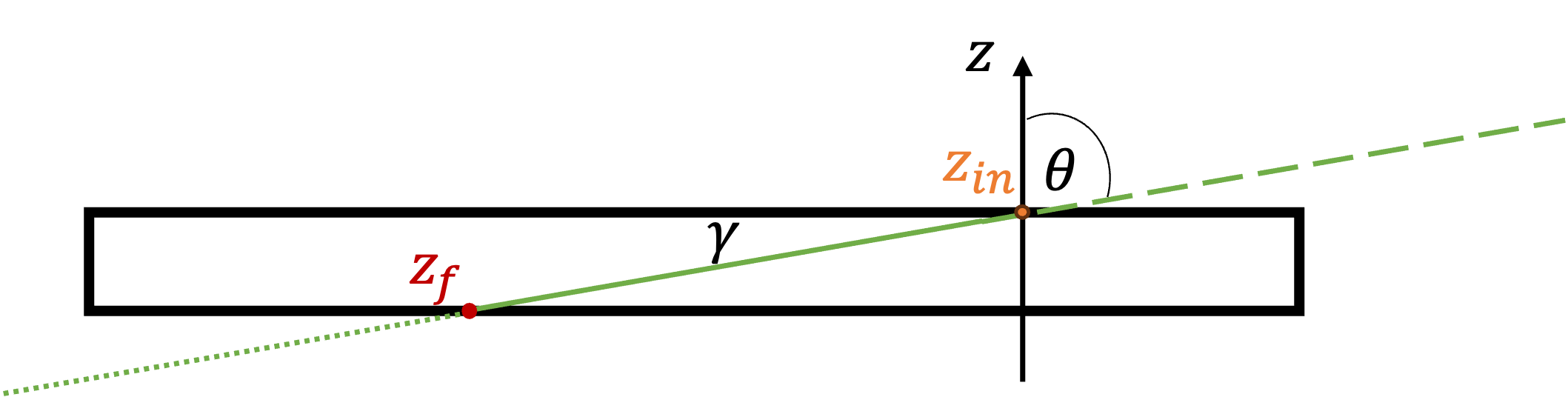}
       \hspace{0cm}(b)
    \end{minipage}
    \caption{(a) Top view of an SSPD (not to scale). (b) Side-cut view of the SSPD along the $(z,\phi, x_{in})$ plane. In green is the track of a non-perpendicularly impinging particle traversing the detector; the dashed green line shows the particle's track before entering the scintillator, the solid green line (from $(x_{in},y_{in},z_{in})$ to $(x_f,y_f,z_f)$) shows the particle track within the scintillator, and the dotted green line shows the particle's possible track if it were to exit the scintillator. Depicted in purple, red, and blue are the angles toward the edges of the top right SiPM at three positions along the particle's track. Photons dispersed to trajectories within these angles reach the top-right SiPM. The particle track is denoted by $\gamma$.}
    \label{fig:2D_model_scintillator_shallow_particle}
\end{figure}

The above-approximated model and equations assume that particles traverse the scintillator perpendicularly. A particle traversing the scintillator non-perpendicularly to the $(x,y)$ plane can be depicted by a superposition of non-vertical light point sources created along the particle track within the scintillator. An accurate representation of the light gathered by each SiPM will be a superposition of the light gathered from each point source. This equates to a superposition of the angles $\alpha$ each light point-source creates with the SiPM along the particle track, as demonstrated in \figref{2D_model_scintillator_shallow_particle}.

We assume that a particle traverses the scintillator from top to bottom, as depicted in \figref{2D_model_scintillator_shallow_particle}, where the particle enters the top part of the scintillator at position $(x_{in},y_{in},z_{in})$ and traverses the scintillator in a non-perpendicular trajectory (represented by the angles $(\theta, \varphi)$) up to a final position $(x_{f},y_{f},z_{f})$. This final position can be the exit position of the particle from the bottom or the side of the scintillator, or from somewhere within the scintillator if the particle comes to a complete stop within the scintillator bulk. In these scenarios, the particle excites the scintillator, which emits photons originating along the particle's track $\gamma$. As each of these excitation points creates a different angle with the SiPMs' edges, the total signal detected by each of the SiPMs is a sum of all the excitation points. Suppose we follow the analytical equations described for the perpendicularly impinging particle model and apply them to this scenario. In that case, it will surely result in an inaccurate estimation of the impinging position and energy deposition.

An accurate representation of the number of photons reaching each SiPM in the case of a non-perpendicularly impinging particle requires integrating the photons excited by the particle along its track within the scintillator.

Using this notation we get
\begin{equation} \label{eq:N_Ph_Formula}
N_{ij} = \frac{\eta_{\text{SiPM}_{ij}} \cdot C_s}{2\pi}\cdot\int_{\gamma} \alpha_{ij}(x,y)\cdot\textrm{LET}(x,y)d\gamma,
\end{equation}
where $\gamma$ is the particle's track inside the scintillator.
In the case where the particle traverses the scintillator and the LET is approximately constant, and since we have $\Delta z = d$,
\eqnref{N_Ph_Formula} reduces to
\begin{eqnarray*}
\label{eq:detector_formulas_full}
N_{ij} &=& \frac{\eta_{\text{SiPM}_{ij}} \cdot C_s \cdot\textrm{LET}}{2\pi}\cdot\int_{\gamma} \alpha_{ij}(x,y)d\gamma . \\
\end{eqnarray*}

Notice that for $\theta =0$, the coordinates $x=x_{in}, y=y_{in}$ are constant which in turn implies that $\alpha_{ij}(x,y) =  \alpha_{ij}$, is constant as well. Hence,
\[
N_{ij} = \frac{\eta_{\text{SiPM}_{ij}} \cdot C_s}{2\pi}\alpha_{ij} \int_0^{d}\mathrm{LET}(t)dt.
\]

Assuming that the LET is approximately constant, we recover \eqnref{detector_formulas_simple}.

\section{Localization and Deposited Energy Estimator} \label{localization_energy_estimator}

Since the geometric equations describing the characteristics of the detector are implicit, we employed a numerical method to estimate the incident location and energy deposited by the detected particle. We begin by developing the algorithm for estimating these parameters assuming a perpendicularly impinging particle. Then, we develop the algorithm for obtaining a more accurate model which considers the particle track.

\subsection{Perpendicularly Impinging Particle Model}

Notice that the four SiPMs measure a signal proportional to the number of photons incident upon each SiPM, 
\[\overrightarrow{N}\equiv\left(\frac{N_{11}}{\eta_{SiPM_{11}}}, \frac{N_{12}}{\eta_{SiPM_{12}}}, \frac{N_{21}}{\eta_{SiPM_{21}}}, \frac{N_{22}}{\eta_{SiPM_{22}}}\right) . \] 
The number of photons
is proportional to the angles $\overrightarrow{\alpha}\equiv(\alpha_{11}, \alpha_{12}, \alpha_{21}, \alpha_{22}).$ 
Assuming a perpendicularly impinging particle event, where the incident angle $\theta=0$:
\begin{equation} \label{eq:PhotonsFormula1}
\overrightarrow{N} = \frac{C_s}{2\pi} \cdot \mathrm{LET} \cdot d \cdot \overrightarrow{\alpha}.
\end{equation}

Given a vector $\overrightarrow{v}$, denote by $\hat{v}$ the normalized vector $\frac{\overrightarrow{v}}{|\overrightarrow{v}|}$.
Notice that by \eqnref{PhotonsFormula1}  $\hat{N} = \hat{\alpha}$. We will use this fact to construct an algorithm that recovers $\overrightarrow{\alpha}$ from $\overrightarrow{N}$.
Consider the map $T:(x,y)\mapsto \hat{\alpha}$. Every tuple of three angles $\alpha_{ij}$ defines $(x,y)$ uniquely, so the normalized vector $\hat{\alpha}$ contains enough information to recover $(x,y)$. In other words, the map $T$ is invertible. Since $\hat{\alpha} = \hat{N}$ we may recover the impinging position using $\overrightarrow{N}$. To find the inverse of $T$, we generated a 2D point grid on the scintillator, with a $1~$ mm spacing between points. For each point on the grid, $T$ was calculated, and stored in a lookup table (BallTree or similar). To calculate the impinging position of a given measurement $\overrightarrow{N}$, it is enough to find the closest element
$\hat{N}$ in the lookup table. The described procedure is listed in Algorithm \ref{algo:alg-FirstOrder}.

\begin{algorithm} 
\caption{Perpendicularly Impinging Particle}
\label{algo:alg-FirstOrder}
\begin{algorithmic}[0]
\small
\State GetAngles(x, y) \Comment{Calculate $\alpha_{ij}$ for a 
given point.}
\State $Res \gets 1 mm$ \Comment{Grid resolution}
\State $ScintSize \gets 70 mm$  \Comment{Scintillator size}
\State $BallTree$ \Comment{Locations to normalized angles table.}

\State
\Procedure{PreProcessing}{}
\State $Grid \gets \{0, 1, \ldots \frac{ScintSize}{Res}\}^2$
\State
\For{each pair $(x, y)$ in $Res\cdot Grid$}
      \State angles = GetAngles(x,y)
        \State BallTree.add$\left(\frac{angles}{|angles|}\right)$
\EndFor
\EndProcedure
\State

\Procedure{GetLocation}{Measurement $\overrightarrow{N}$}
    \State $(x,y) = \textrm{BallTree.find}\left(\frac{\overrightarrow{N}}{|\overrightarrow{N}|}\right)$
    \State return x,y
\EndProcedure
\end{algorithmic}
\end{algorithm}

\subsection{Non-Perpendicularly Impinging Particle Model}

To generalize the algorithm for non-perpendicularly impinging particles, we consider a model of a pair of SSPDs stacked vertically. This way, each particle that traverses both detectors provides eight measurements, which suffice to recover its track and deposited energy.

Our algorithm resembles the specific case of a perpendicularly impinging particle approximation described in the previous section. Given two points $ p_1 = (x_1, y_1) $ and $ p_2 = (x_2, y_2) $ in scintillators 1 and 2, respectively, we can recover the particle's track, and calculate $ N_{ij} $ from  \eqnref{detector_formulas_full}, up to a constant. Given a particle's track $ \gamma $ in scintillator $ k $, denote:
\[
A^{(k)}_{ij} = \int_{\gamma} \alpha_{ij}(x,y)d\gamma,
\]
where $\alpha_{ij}(x,y)$ are the angles defined in \eqnref{detector_formulas_full}.

The map $ T:(p_1, p_2) \mapsto (\hat{A}^{(1)}, \hat{A}^{(2)}) $ is invertible, so given two vectors $ (\overrightarrow{N}^{(1)}, \overrightarrow{N}^{(2)}) $, we may recover the tuple $ (p_1, p_2) $.
This is done similarly to the previous algorithm: A two-dimensional grid with a resolution of $1$~mm on the surface of each scintillator was generated. Given two measurements $ (\overrightarrow{N}^{(1)}, \overrightarrow{N}^{(2)}) $, we look for the closest element 
$ (\hat{N}^{(1)}, \hat{N}^{(2)}) $ in our lookup table, which provides us with the approximation of points $ (p_1, p_2) $ that define the particle's track uniquely. Having the particle's track allows an approximation of its LET in each scintillator in the case where it was approximately constant. We summarize the described procedure in Algorithm \ref{algo:alg-SecondOrder}.
\begin{algorithm} 
\caption{Non-perpendicularly impinging particle}
\label{algo:alg-SecondOrder}
\begin{algorithmic}[0]
\small
\State GetAngles($x_1, y_1, x_2, y_2$) \Comment{Calculate $A^{(k)}_{ij}$ for every $(x,y)$ impinging particle locations in every two consecutive scintfilators.}\\

\State $Res \gets 1 mm$ \Comment{Grid resolution}
\State $ScintSize \gets 70 mm$  \Comment{Scintillator size}
\State $BallTree$ \Comment{Locations to normalized angles table.}

\State
\Procedure{PreProcessing}{}

\State $Grid \gets \{0, 1, \ldots \frac{ScintSize}{Res}\}^4$
\State 
\For{each tuple $(x_1, y_1, x_2, y_2)$ in $Res\cdot Grid$}
     \State $(\overrightarrow{A}^{1}, \overrightarrow{A}^{2})$ = GetAngles$(x_1,y_1, x_2, y_2)$ 
   \State BallTree$[x_1, y_1, x_2, y_2]$ = $\left(\frac{\overrightarrow{A}^{1}}{|\overrightarrow{A}^{1}|}, \frac{\overrightarrow{A}^{2}}{|\overrightarrow{A}^{2}|}\right)$
 \EndFor

\EndProcedure
\State

\Procedure{GetLocation}{Measurements $\overrightarrow{N}_1, \overrightarrow{N}_2$}
    \State $(x_1,y_1, x_2, y_2) = \textrm{BallTree.find}\left(\frac{\overrightarrow{N}_1}{|\overrightarrow{N}_1|}, \frac{\overrightarrow{N}_2}{|\overrightarrow{N}_2|}\right)$
    \State return $x_1,y_1, x_2, y_2$
\EndProcedure
\end{algorithmic}
\end{algorithm}

\section{Detector Implementation} \label{detector_implementation}

In this section, we provide a detailed account of these detectors' design and production, followed by comprehensive laboratory calibration and testing.

\subsection{Design and Manufacturing}

\begin{figure}[htbp]
    \centering
    \begin{minipage}[t]{0.7\textwidth}
        \centering
        \includegraphics[width=\linewidth]{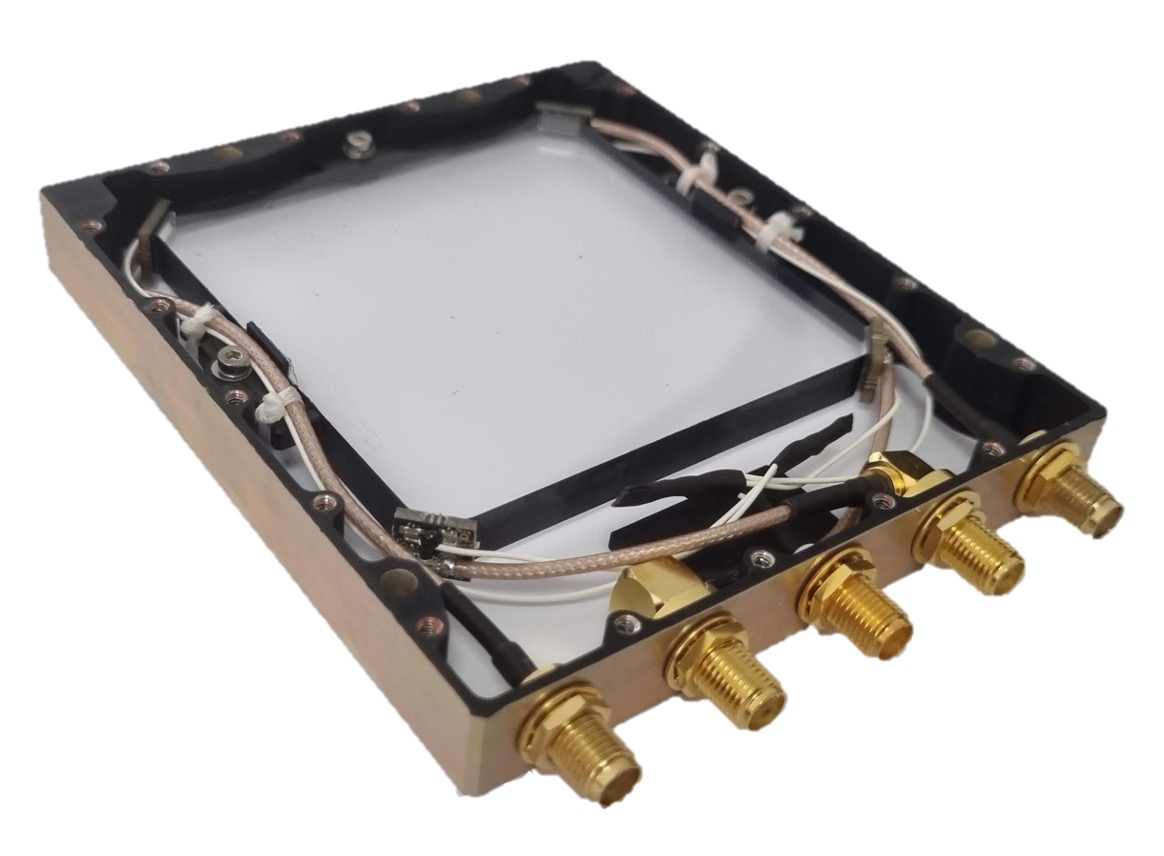}
        \centering (a)
    \end{minipage}\\
    \hspace{0.25cm}\begin{minipage}[t]{0.7\textwidth}
        \centering
        \includegraphics[width=\linewidth]{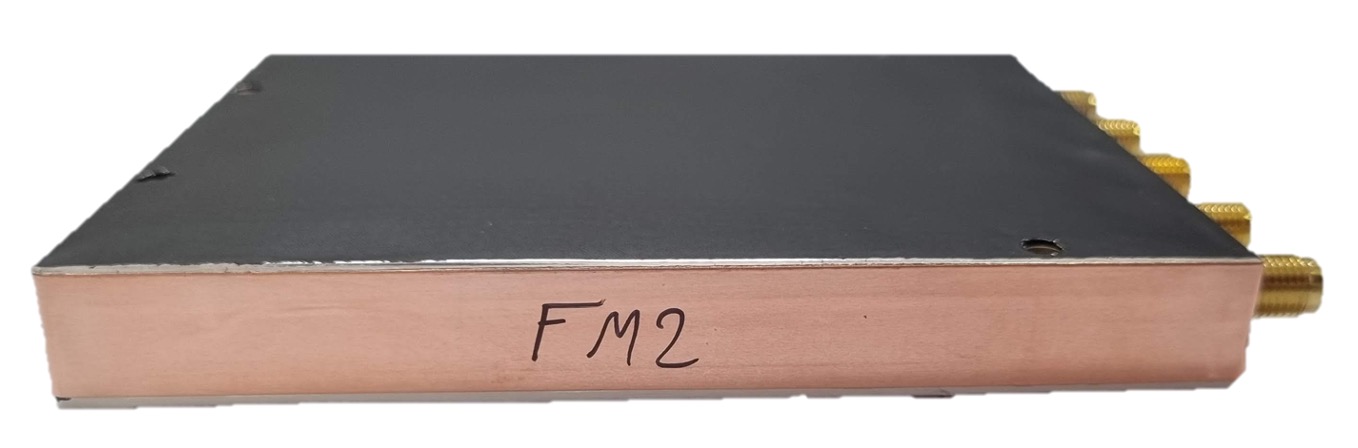}\label{fig:figure6}
        (b)
    \end{minipage}
    \caption{(a) Open-covered COTS-Capsule SSPD, devoid of top and bottom covers. Noteworthy in the image is the black aluminum frame housing and the four SensL FJ-60035 SiPMs, securely glued to the EJ-200 scintillator. The blackened edges of the scintillator are also discernible. Subsequently, this detector is enveloped by Thorlabs BKF12 matte black 70~$\mu$m aluminum foil.
    (b) Fully enclosed flight module of the SSPD which was eventually launched aboard the COTS-Capsule spaceborne experiment~\cite{COTS-Capsule-System}}
    \label{fig:real_scintillator}
\end{figure}

The SSPDs are custom-crafted specifically for the COTS-Capsule space experiment that was launched to the ISS~\cite{COTS-Capsule-System}. Our design is centered around a milled and polished square $70 \times 70 \times 6.7~$ mm$^3$ prism-shaped scintillator, utilizing Eljen's EJ-200 polyvinyl toluene polymer scintillator~\cite{eljen_scintillator} as the core material. To optimize the detector's performance, the scintillator's corners are truncated to create $6.7 \times 6.7~$ mm$^2$ "truncated corner faces" that are illustrated in \figref{3D_model_scintillator_detector} and \figref{real_scintillator} adjacent to the green electronic boards. 

Equipped with four electronic boards, each housing a SensL FJ-60035 SiPM sensor, the scintillator sensors are optically coupled to the truncated corner faces using Eljen's optical cement (EJ-500). Achieving precise alignment of the SiPM sensors to the truncated corner faces requires a carefully crafted positioning jig. These SiPM sensors' electronic boards are then connected to panel-mounted SMA connectors, supplying a bias voltage to the sensors and transmitting signals from the sensors toward a readout and acquisition module.

To minimize stray light, the side faces of the scintillator are coated with Eljen's EJ-510B light-absorbing black paint. Polyacetal copolymer interface brackets are affixed to the scintillator's side faces using 3M 966 double-sided tape. Once cured, this process results in a sub-assembly seamlessly integrated into the detector frame, securely holding the scintillator in place without interfering with the top and bottom optical surfaces.

The detectors were enclosed in a light-tight, internally light-absorbing metal enclosures made from CNC-milled aluminum frames. The internal surfaces of these frames were coated with an inorganic black finish and carefully covered with Thorlabs BKF12 black aluminum foil, measuring $70~\mu$m in thickness. This foil is crucial for absorbing stray light within SSPDs and improving measurement accuracy. A template is used to ensure precise foil shaping and the black epoxy covering the perimeter of the foil is removed to maintain electrical conductivity along the frame assembly's outer edges. Copper tape is utilized to secure the aluminum foil to the metal frame, effectively reducing Electro-Magnetic Interference (EMI), as illustrated in \figref{real_scintillator}. These design considerations are instrumental in minimizing EMI within the detector, ensuring accurate position and energy deposition estimations.

\subsection{Testing and Calibration}

We conducted several test campaigns for detector characterization and calibration, including calibration with a 2D scanning Ru$^{64+}$ $\beta^-$ radioactive source and calibration with secondary cosmic ray muons and small coincidence detectors placed above and below the SSPDs under test. The most efficient and accurate calibration method was achieved by mutually calibrating a set of SSPDs in the laboratory utilizing the COTS-Capsule hodoscope~\cite{COTS-Capsule-System}.

\begin{figure}[htbp]
\centering
\includegraphics[width=0.7\textwidth]{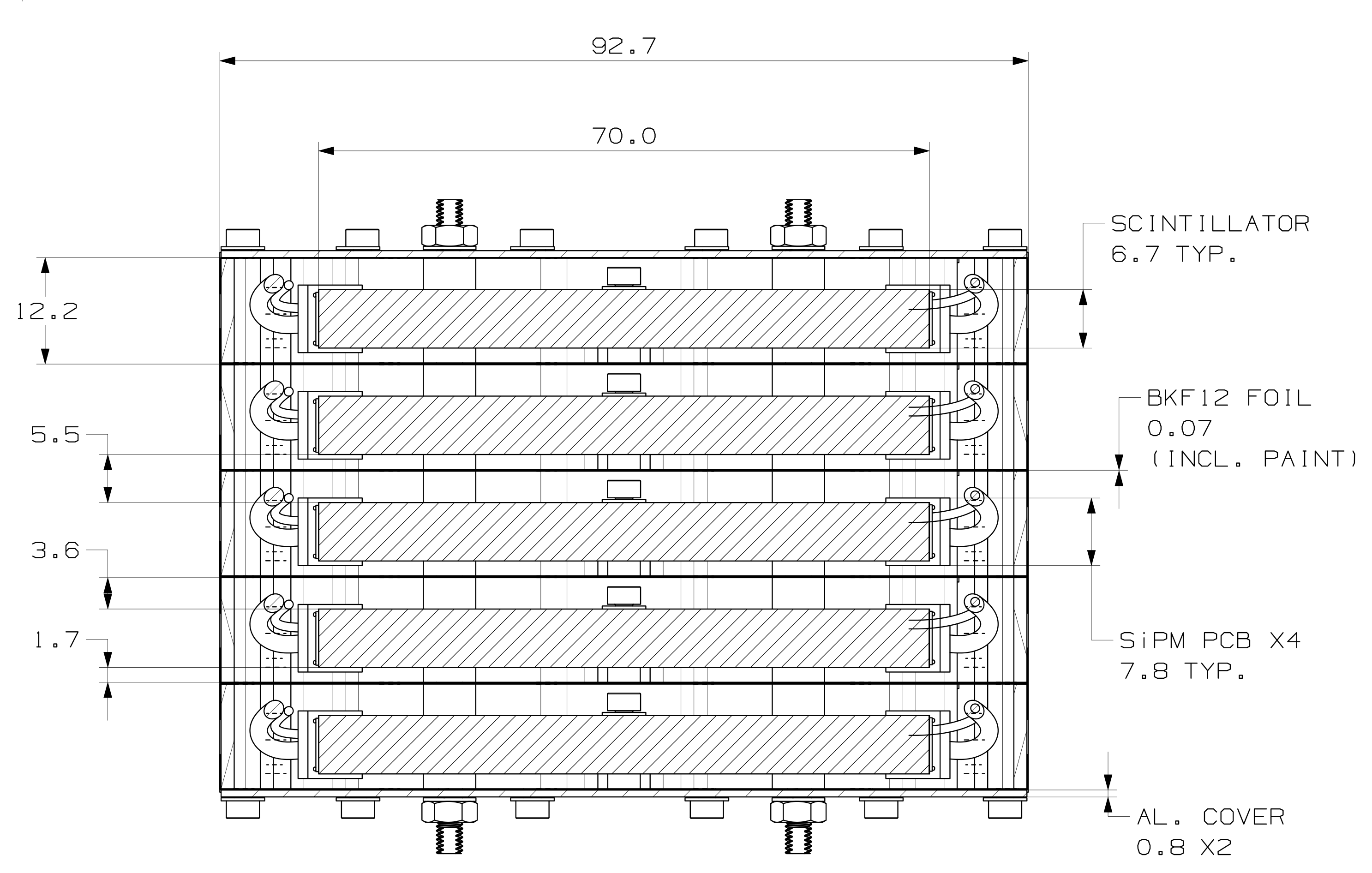}\label{fig:figure5}
\caption{Particle detector hodoscope design for the COTS-Capsule spaceborne experiment~\cite{COTS-Capsule-System}. The five vertically aligned SSPDs are marked with a diagonal line filling.}
  \label{fig:telescope_sidecut_view}
\end{figure}

Although each SSPD functions independently and is self-contained, the detectors were designed to be stacked coaxially, one on top of the other in a particle hodoscope configuration as shown in \figref{telescope_sidecut_view}. By stacking several (at least three) particle detectors one on top of the other, we calibrated the detectors in the laboratory using muons originating from natural secondary cosmic rays.

Two base physical assumptions promoted our mathematical calibration approach: 
\begin{enumerate}
    \item The flux of cosmic rays is roughly invariant to rotation around the Zenith~\cite{Workman:2022ynf}.
    \item Muons that traverse the entire length of the hodoscope do so in nearly straight paths~\cite{Workman:2022ynf}.
\end{enumerate}
 
We perform linear regression to test the calculated uncertainty. 

For a particle entering the SSPD at its top surface and exiting at the bottom surface, the number of photons arriving at each SiPM, based on the first approximation of our physical model for the special case of a perpendicularly impinging particle, is:

\begin{equation}
    N^{(k)}_{ij} = C^{(k)}_{ij}\cdot \frac{C_s}{2\pi} \cdot d \cdot  {\rm{LET}} \cdot \alpha_{ij}^{(k)} ,
\end{equation}

where:
\begin{itemize}
\item Scintillator $k$ is the $k^{th}$ scintillator down from the top of the hodoscope.
\item SiPM$_{ij}^{(k)}$ is the detector whose index is $ij$, attached to the $k^{th}$ scintillator.
\item $N^{(k)}_{{ij}}$ is the number of photons arriving at SiPM$_{ij}^{(k)}$. Assuming the SiPM is in its linear response region the number of photons is proportional to the signal gathered from the SiPM.
\item $C^{(k)}_{ij}$ is a constant depending on the SiPM$_{ij}^{(k)}$ efficiency (unknown).
\item LET$^{(k)}$ is the particle's Linear Energy Transfer while traversing scintillator $(k)$.
\item $\alpha^{(k)}_{ij}$ is the angle between the edges of SiPM$_{ij}^{(k)}$ and the particle's impinging position (see \figref{2D_model_scintillator_detector_angles_to_sensors}).
\end{itemize}
Utilizing the aforementioned model provides us with adequate data for calibrating the hodoscope, even in the absence of precise knowledge of the parameters $C^{(k)}_{ij}$. Since this calibration algorithm is applied individually to each scintillator, we can assume that $k$ is a given parameter and thus omit the upper indices.

Following assumption (1), it is reasonable to assume that the incident particle's impinging location $(x,y)$ exhibits rotational symmetry with respect to the $z$-axis. This leads to the following expectation relation: $\mathbb{E}[\alpha_{i'j'}] = \mathbb{E}[\alpha_{ij}]$, or equivalently, 
\[
0 = \mathbb{E}\left[\log\frac{\alpha_{ij}}{\alpha_{i'j'}}\right] = \log{\frac{C_{i'j'}}{C_{ij}}} + \mathbb{E}\left[\log\frac{N_{ij}}{N_{i'j'}}\right].
\]

The above formula yields 
\[
\frac{C_{ij}}{C_{i'j'}} = \exp\left(\mathbb{E}\left[\log\frac{N_{{ij}}}{N_{{i'j'}}}\right]\right).
\]
Using $\frac{C_{ij}}{C_{i'j'}}$, we can recover the particle's location using both algorithms described in Sec.~\ref{localization_energy_estimator}. The key insight is that without knowledge of $C_{ij}$, we cannot derive the angles $\vec{\alpha}$ that define the incident point $(x_k,y_k)$. However, $(x_k,y_k)$ can be uniquely determined knowing the vector $\hat{\alpha}$.

This vector can be calculated using the measured $\frac{C_{ij}}{C_{i'j'}}~\cdot 
~N_{ij}$, which is proportional to $\alpha_{ij}$ with a ratio independent of the indices $i$ and $j$. Subsequently, we apply the chosen algorithm to each measurement to reconstruct the impinging location and, up to a system-specific scaling constant, the LET. 

\section{Instrument Modelling and GEANT4 Simulations} \label{instrument_modelling_GEANT4}

To test our models, we created a GEANT4 simulation of two SSPDs stacked vertically.

SSPDs were designed and manufactured for spaceborne operations, primarily to detect and characterize heavy-ion impinging particles. Given that the detector was primarily designed to measure high LET ions in space, we simulated high-energy oxygen-16 ions with an energy of $18~$GeV. Nevertheless, since available particles in the laboratory are secondary cosmic particles, mainly muons, we also performed GEANT4 simulations of muon events to compare laboratory results with simulated results.

The next step involved generating simulations of the mentioned particles, with a particle gun located at a random location and directed towards a random location within the contour of the top detector.

To apply the physical model algorithms, we filtered only particles that traversed both scintillators, amounting to roughly $3 \times 10^4$ simulation events per particle type.

It is important to recognize several factors that differentiate the "real-world" detector performance from the theoretical geometrical model described in Sec.~\ref{Design and Physical Model of the SSPD}. The simulation accurately models the physical interaction of particles traversing through matter such as nuclear interactions, energizing delta electrons, small deviations from a straight trajectory, etc. Other factors that were not modeled include:

\begin{itemize}
\item Non-ideal optical surfaces, specifically on the top and bottom surfaces of the scintillator, causing light scattering and degradation of total internal reflection.
\item Non-ideal geometry, accounting for manufacturing tolerances.
\item Non-ideal absorption characteristics of the black paint on the scintillator's edges, the black anodize on the enclosure, and the black epoxy paint of the BKF-12 aluminum foil.
\item The reflection from the SiPMs' surface.
\item Non-ideal SiPM sensor characteristics, including noise, crosstalk, and saturation.
\end{itemize}

\section{Results}

In this section, we assess the performance of the physical-model algorithms using data obtained from GEANT4 simulations and laboratory experiments conducted with the manufactured detectors.

\subsection{GEANT4 Simulation Results}

The GEANT4 simulations, discussed in Sec.~\ref{instrument_modelling_GEANT4}, serve as a theoretical benchmark for interpreting subsequent laboratory results. We rigorously tested the detectors' calibration and the accuracy of position and energy estimation algorithms by comparing them with the 'ground-truth' simulated data. It's anticipated that the performance of the GEANT4 algorithm may slightly surpass that measured in the laboratory due to deviations between the GEANT4 model and the "real-world" detector as depicted in Section \ref{instrument_modelling_GEANT4}.

\subsubsection{Perpendicularly Impinging Particle Approximation}

We applied Algorithm \ref{algo:alg-FirstOrder} to simulated high-energy oxygen particle events impacting the scintillator at randomly chosen points and directions, i.e. non-perpendicularly impinging particles, using GEANT4. We estimated particle impinging position and LET. The root mean square error of the position estimation was $2.9~$ mm, with a LET relative uncertainty of $10\%$.

\subsubsection{Non-Perpendicularly Impinging Particle Approximation}

\begin{figure}[htbp]
\centering
\includegraphics[width=0.7\textwidth]{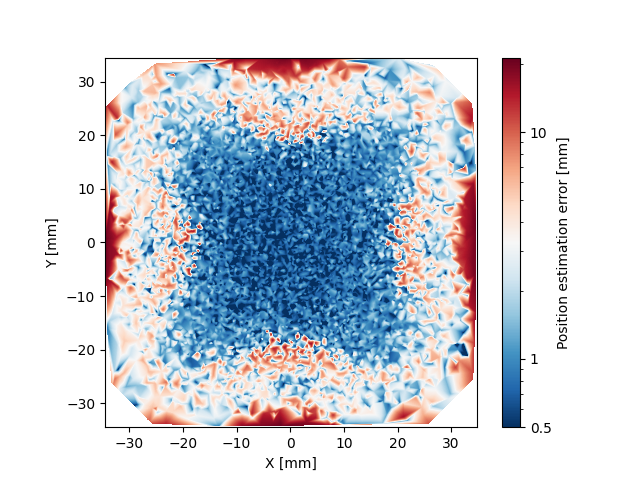}
\caption{Heat map illustrating the estimation errors of impinging particle positions according to the non-perpendicularly impinging particle algorithm. The dataset consists of $3 \times 10^4$ oxygen ion events simulated in GEANT4, with a standard deviation of $2.3~$mm.}
\label{fig:location_error_3d_enter}
\end{figure}

\begin{figure}[htbp]
\centering
\includegraphics[width=0.7\textwidth]{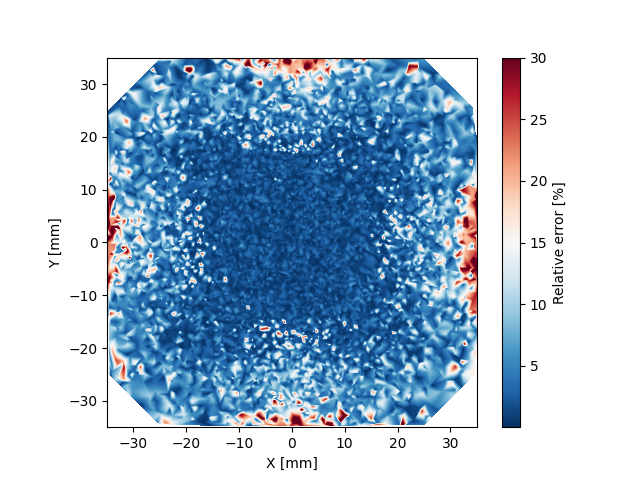}
\caption{Heat map illustrating the relative uncertainties of impinging particle LET according to the non-perpendicularly impinging particle algorithm. The dataset comprises $3 \times 10^4$ oxygen ion events simulated in GEANT4. The LET relative uncertainty is $5$$\%$.}
\label{fig:oxygen_5scints}
\end{figure}

Utilizing the same simulation results presented in the previous section, we proceeded to apply the non-perpendicularly impinging particle Algorithm \ref{algo:alg-SecondOrder}. The root mean square error of the location estimation was $2.3$~mm, as depicted in \figref{location_error_3d_enter}. Additionally, the LET relative uncertainty was $5\%$, as illustrated in \figref{oxygen_5scints}.

These results demonstrate an improvement over the perpendicularly impinging particle algorithm, as the non-perpendicular algorithm accounts for the angular track of the particle through the scintillator and the associated nonlinear effects. Notably, the detector's performance is improved for particles impinging at the center of its sensitive area. However, near the detector's borders, accuracy diminishes due to increased Geometric Dilution of Precision (GDOP). The 'texture' observed in the estimation uncertainty is attributed to the inherent instability of the numerical solution when applied to the implicit analytical formulas of the physical model.

\subsection{Laboratory Detector Calibration Results}

The five-layered hodoscope commenced data collection from secondary cosmic particle events for over a year. These events primarily consisted of muon particle events~\cite{Workman:2022ynf}. During this period, the hodoscope accumulated data from more than $3.2 \times 10^7$ events, aligning with the secondary cosmic ray flux in the laboratory and the geometry of the hodoscope~\cite{Workman:2022ynf}.

For calibration purposes, we exclusively considered events where all $20$ SiPMs detected signals within their dynamic range. Subsequently, employing the methods and formulas outlined in the "Testing and Calibration" section, we optimized and calibrated the individual SiPM coefficients. Additionally, we estimated the positions where particles impinged the horoscope and the energies these particles deposited using data collected from laboratory measurements.

\subsubsection{Perpendicularly Impinging Particle Approximation}

Using the perpendicularly impinging particle Algorithm \ref{algo:alg-FirstOrder}, we computed the estimated impinging particle positions in each scintillator. For each event, we calculated the best-fitting track using standard linear regression based on the estimated impinging positions across all detectors. 

Let $(x_i, y_i)$ denote the estimated particle's position in scintillator $i$, and $(\hat{x}_i, \hat{y}_i)$ denote the predicted location using linear regression. This fitted track served as the 'average ground truth'. For each event, we determined the estimation uncertainty, defined as the difference between each detector's impinging particle position estimation and the fitted track, i.e., $(\Delta x_i, \Delta y_i) = (x_i - \hat{x}_i, y_i - \hat{y}_i)$. The resulting standard deviation of the particle position estimation using linear regression is $\sigma_{x_3}=6.2~$ mm and $\sigma_{y_3}=6.1~$ mm (for the third detector).

\subsubsection{Non-Perpendicularly Impinging Particle Approximation}

To assess the non-perpendicularly impinging particle algorithm, we applied it to each pair of consecutive scintillators, resulting in estimates for the particle's track at eight scintillator surfaces along the hodoscope - encompassing all exit surfaces of all detectors except for the last one, and all entry surfaces of all detectors except for the first one. Subsequently, we employed linear regression to estimate the predicted impinging position. Similar to the perpendicularly impinging particle algorithm, we calculated the deviation vector $(\Delta x_i, \Delta y_i) = (x_i - \hat{x}_i, y_i - \hat{y}_i)$ for each point.

\begin{figure*}
	\centering
		\includegraphics[width=\textwidth]{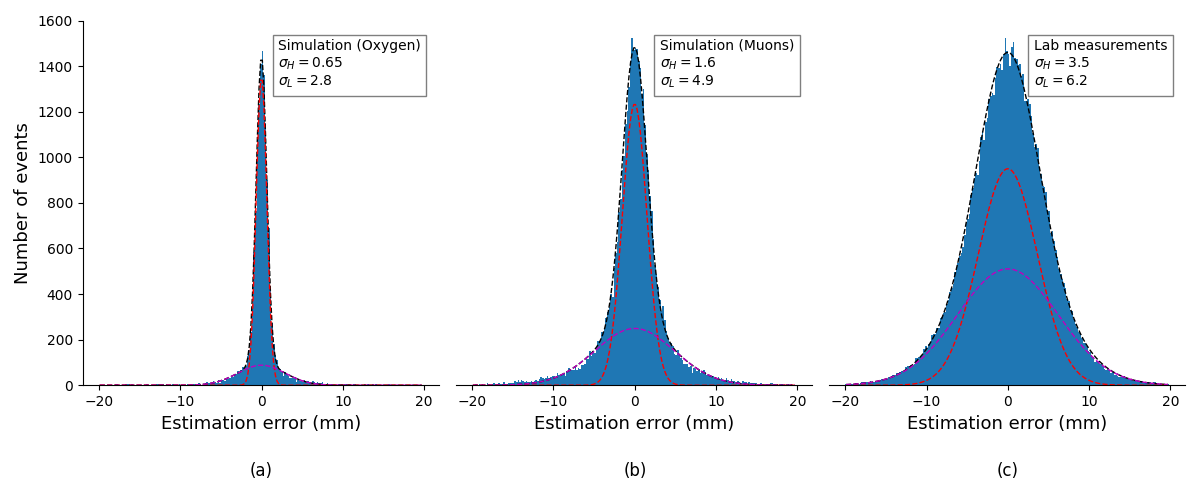}\label{fig:gaussians}
    \caption{Histograms of impinging particle position estimation error according to Algorithm \ref{algo:alg-SecondOrder}. (a) Histogram depicting the GEANT4 simulation results for high-energy ($18~$GeV) oxygen ions. (b) Histogram illustrating the GEANT4 simulation outcomes for high-energy ($4~$GeV) muons. (c) Histogram showing the secondary cosmic particle measurements by the laboratory hodoscope. The dashed black line represents a two-Gaussian sum fit to the data. The fit is the sum of the red dashed line Gaussian fit and the purple dashed line Gaussian fit. 
    } 
  \label{fig:delta_xy_dist_3d}
\end{figure*}

The discrepancy between the algorithm's estimated impinging particle position per detector-pair and the linear regression of all detector-pairs (considered the 'average ground truth') is depicted in the histograms shown in 
\figref{delta_xy_dist_3d}. These histograms illustrate the impinging particle position estimation error according to the non-perpendicularly impinging particle algorithm. The left graph represents the simulated high-energy ($18~$GeV) oxygen ions, the middle graph depicts high-energy ($4~$GeV) muons, and the right graph illustrates secondary cosmic particles, mostly Minimum Ionizing Particles (MIP) muons, measured by the hodoscope in the laboratory.

The standard deviation of the position estimation error using linear regression for lab-measured secondary cosmic particles is $\sigma_{x_3}=5.1~$mm and $\sigma_{y_3}=5.05~$mm (for the third detector), representing a $20\%$ improvement compared to the perpendicularly impinging particle algorithm.

As expected, the histograms in  \figref{delta_xy_dist_3d} show that the GEANT4 simulations exhibit slightly better impinging particle position estimation error than the measurements gathered by the hodoscope in the laboratory, with the oxygen ion GEANT4 simulations showing considerably better performance. These results are summarized in \tableref{algorithm_results_summary}.

To properly fit the data, a double Gaussian distribution is required. This is attributed to the variability in the detector's and algorithm's position error within the detector's sensitive area. \figref{location_error_3d_enter} shows that the detector's position estimation error in the middle of the detector's sensitive area is better represented by the thinner Gaussian distribution. In contrast, the wider Gaussian distribution better represents the position estimation error near the boundaries of the scintillator.

\begin{figure*}
	\centering
		\includegraphics[width=\textwidth]{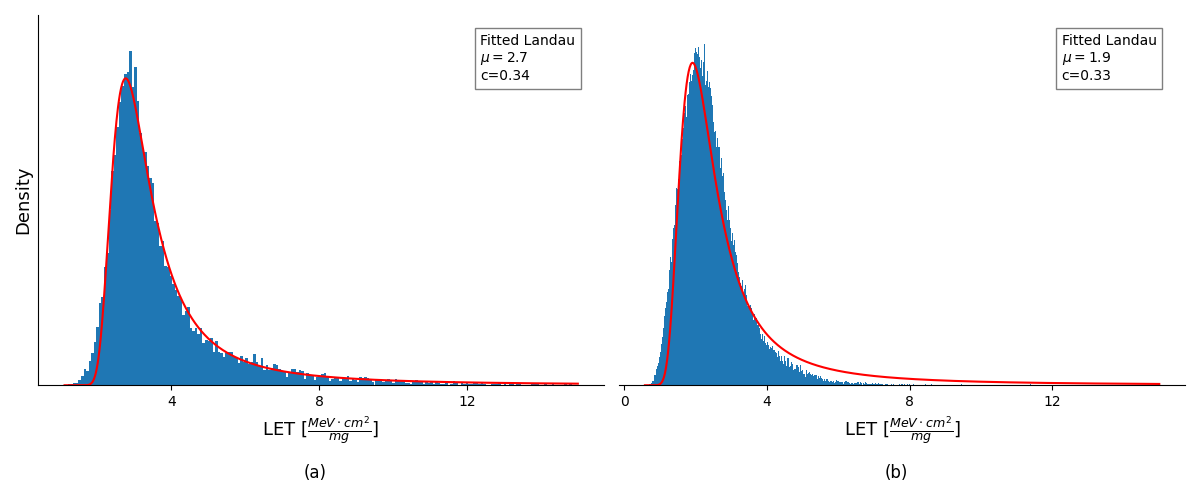}\label{fig:landaus}
    \caption{Histograms depicting the estimated-LET distribution for (a) GEANT4 simulated muons with energies of $4~$GeV, and (b) secondary cosmic particles measured in the laboratory, each fitted with a Landau distribution curve. The LET was estimated by applying Algorithm \ref{algo:alg-SecondOrder}. Notice that the distribution on the right has a clipped tail. This is due to filtering of results which are below the saturation threshold of the system.}
  \label{fig:3d_let}
\end{figure*}

The algorithms also provide an estimation of the deposited energy within each detector as well as the impinging particles' tracks. By dividing the deposited energy by the particle's track length within the scintillator, it is possible to estimate each impinging particle's LET. 
\figref{3d_let} provides histograms depicting estimates of simulated and measured impinging particles' LET.

While traversing through matter, particles deposit some of their energy, thereby losing speed and changing their LET. We carefully simulated only high-energy MIPs to ensure minimal change in LET during passage through the thin detectors. Similarly, for laboratory measurements, we reduced the dataset to particles detected by all five SSPDs and estimated the LET for only the first SSPD. This increases the ratio between MIP muons events relative to non-MIP muons events and ensures minimal LET change within the first scintillator. 

\figref{oxygen_5scints} presents a heat map of the LET relative uncertainties of simulated events based on the non-perpendicularly impinging particle algorithm. \figref{3d_let} left shows the LET relative uncertainty for simulated high-energy ($4~$GeV) muons, while the right graph provides the LET relative uncertainty in lab-measured secondary cosmic particles.

Both histograms resemble a landau distribution, expected for a MIP traversing a thin scintillator. The absolute LET value for laboratory measurements was calibrated according to the most probable LET value for muons in polyvinyl toluene~\cite{Workman:2022ynf}.

It is discernible that the histogram provided for the laboratory measurements falls short of the Landau distribution along its trailing edge (higher LET). 
This phenomenon results from a limitation imposed on the lab-gathered dataset, which omitted events where one or more of the twenty SiPMs were saturated. 
Such exclusion criterion eliminates high-LET particles. 
This observation was confirmed by replicating the phenomenon through the same exclusion criteria applied to the GEANT4 simulated muon events, thereby reproducing this effect.

\section{Discussion}

In this paper, a specific SSPD design was presented, featuring a rectangular prism scintillator of dimensions $70\times 70\times 6.7~$ mm$^3$ and four SiPM sensors on its truncated corner faces. This design can be scaled with variations in dimensions or sensor numbers while maintaining fundamental principles. 

The number of SiPM sensors determines the detector's ability to estimate the free parameters. Five parameters are required for particles with consistent LET and a non-perpendicular incident angle: particle impinging position in both $x$ and $y$ directions, total energy deposition within the scintillator, and the incident angles $\theta$ and $\phi$. A single SSPD with at least five SiPM sensors, such as pentagonal or hexagonal prisms, can provide a comprehensive parameter suite, with additional sensors aiding in parameter refinement and error reduction. Complex scenarios, such as particles stopping within the detector mid-track, may require additional sensors and more intricate physical models.

The detectors' scalability extends to all dimensions, allowing adjustments to effective area and parameter estimation precision. This enhances the SSPDs applicability across diverse applications. Note that to compensate for changing the SSPD's dimensions adjustments in SiPM area might be required.

\section{Summary and Conclusions}

We introduce a novel SSPD detector that offers advantages over traditional methods. The detector, designed and manufactured for the COTS-Capsule spaceborne experiment~\cite{COTS-Capsule-System}, exhibits high accuracy, low implementation complexity, reduced power and thermal dissipation, low cost, and scalability. 

The detector's performance was evaluated through physical modeling and analytic formulation, GEANT4 simulations, and laboratory tests. 

The design of the SSPD relies on a single element of scintillating material and a sparse number of SiPM sensors and electronic readout channels. This design facilitates a 2D intensity-based triangulation method, distinguishing it from traditional time-based triangulation methods. 

Our findings demonstrate localization accuracy for high-energy oxygen ions of approximately $2.3~$mm and relative uncertainties in impinging particle LET estimation around $5\%$ across the total available detector area. Significantly improved results are obtained when limiting the active detection area to a smaller region around the detector's center.

\begin{table*}
    \centering
    \begin{tabular}{|l|p{2.5cm}|c|c|c|c|} \hline 
        Setup&Particle type and energy& \multicolumn{2}{p{3.8cm}|}{Perpendicularly \ \ \ \ \ \ \ \ \ \ \ \ \ \ \ \ \ \ \ \ \ 
 impinging particle algorithm} & \multicolumn{2}{p{3.8cm}|}{Non-perpendicularly impinging particle algorithm}\\ \hline 
        && $\sigma_{x_3}$ $[mm]$ &$\nicefrac{\sigma_{\text{LET}}}{\text{LET}}$ $[\%]$ & $\sigma_{x_3}$ $[mm]$ & $\nicefrac{\sigma_{\text{LET}}}{\text{LET}}$ $[\%]$\\ \hline 
        GEANT4
& Oxygen ions, \ \ \ \ \ \ \ \ \ \ \ \ 
18~GeV&2.8 & 10&2.3 &  5\\ \hline 

        GEANT4
& Muons, \ \ \ \ \ \  \ \ \ \ \ \ \ \ \ \ \ \ \ \ \ \ \ \     
4~GeV&5 & 22& 3.7 & 12\\ \hline

        Laboratory & Muons,\ \ \ \ \ \ \ \ \ \ \ \ \ \ \ \ \ \ \ \ \ \ \ \ \ \ \ \   
sea level
secondary cosmic particles &6.2 & N/A& 5.1 & N/A\\ \hline
    \end{tabular}
    \caption{Performance summary of algorithms across various particle types and energies: simulations vs. laboratory results. The results are shown for the third SSPD and only for the x-axis, as the results for the y-axis are similar.}
    \label{tab:algorithm_results_summary}
\end{table*}

\section{Acknowledgements}

We extend our heartfelt gratitude to all individuals and organizations whose contributions were instrumental to the success of the COTS-Capsule mission to the International Space Station. Eljen, for their exceptional craftsmanship, dedication, and meticulous attention to detail. CAEN for their invaluable support with the DT5702. The Soreq Nuclear Research Center provided essential guidance and technical assistance through their space environment team. Our colleagues at Tel Aviv University, including the Physics Machine Shop, for their guidance and support throughout the project. Special thanks to Meny Raviv Moshe, Dr. Igor Zolkin, and Prof. Yan Benhammou. Etzion acknowledges the support of the Canada First Research
Excellence Fund. We thank the Israel Aerospace Industries for their expertise, insightful feedback, and constant support.

\bibliography{SSPD}
\bibliographystyle{unsrt}

\end{document}